\documentclass{article}

\usepackage{arxiv}

\usepackage[utf8]{inputenc} 
\usepackage[T1]{fontenc}    
\usepackage{hyperref}       
\usepackage{url}            
\usepackage{booktabs}       
\usepackage{amsfonts}       
\usepackage{amsmath}
\usepackage{bm}
\usepackage{cleveref}
\usepackage{nicefrac}       
\usepackage{microtype}      
\usepackage{lipsum}
\usepackage{graphicx}
\usepackage[
    backend=biber,
    style=numeric,
    sorting=nty,
    maxnames=8,
    minnames=3,
    maxcitenames=2,
    mincitenames=1,
    abbreviate=false,
    giveninits=true,
    date=year
]{biblatex}
\graphicspath{ {./fig/} }

\bibliography{bib}

\newcommand{\ftheta}{\bm{\theta}}

\title{Transport Map Coupling Filter for State-Parameter Estimation}

\begin{document}

\maketitle

\begin{center}
	\vspace*{0.5cm}
	\normalsize{Jan Grashorn$^{a,b,*}$, Matteo Broggi$^{a}$, Ludovic Chamoin$^{b,c}$, Michael Beer$^{a,b,d,e}$}\\
	\vspace*{0.5cm}
	\footnotesize{$^a$Institute for Risk and Reliability, Leibniz Universität Hannover, Hannover, Germany}\\
	\footnotesize{$^b$International Research Training Group 2657, Leibniz Universität Hannover, Hannover, Germany}\\
	\footnotesize{$^c$Université Paris-Saclay, CentraleSupelec, ENS Paris-Saclay, CNRS, LMPS - Laboratoire de Mécanique Paris-Saclay, Gif-sur-Yvette, France}\\
	\footnotesize{$^d$Institute for Risk and Uncertainty, University of Liverpool, Liverpool, United Kingdom}\\
	\footnotesize{$^e$International Joint Research Center for Resilient Infrastructure \& International Joint Research Center for Engineering Reliability and Stochastic Mechanics, Tongji University, Shanghai, China}\\
	\vspace*{0.2cm}
	\footnotesize{$^*$Corresponding author, grashorn@irz.uni-hannover.de}\\
	\vspace*{0.5cm}
	Accepted in \textit{Advances in Reliability, Safety and Security ESREL 2024 Contributions} on 03.04.2024
\end{center}

\vspace{.2cm}

\begin{abstract}
Many dynamical systems are subjected to stochastic influences, such as random excitations, noise and unmodeled behavior. Tracking the system’s state and parameters based on a physical model is a common task for which usually filtering algorithms, e.g. Kalman filters and their non-linear extensions, are used. Many of these filters however use assumptions on the transition probabilities or the covariance model which can lead to inaccuracies in non-linear systems. We will show here the application of a stochastic coupling filter that is able to approximate arbitrary transition densities under non-Gaussian noise. The filter is based on so-called transport maps which couple the approximation densities to a user-chosen reference density and thus allow for straight-forward sampling and evaluation of probabilities.
\end{abstract}


\section{Introduction}
Dynamical systems are at the heart of many engineering disciplines. Knowledge about the system's internal parameters is important in these to ensure safe operation conditions, to predict future behavior or to gain insight on the system's condition, i.e. for damage detection and health monitoring. This knowledge is obtained or updated through measurements from some system components, however, in many cases it is not possible to directly measure the system's state. Moreover, the measurements are polluted by inherent sensor noise which needs to be filtered out. In the past, many algorithms have been developed towards this goal. The most prominent is the Kalman filter, which, due to its optimality in linear systems subjected to Gaussian noise, still is rightfully used in many applications. For non-linear systems there have been many endeavors towards extending the Kalman filter's capacity, however, these often rely on the Gaussian noise assumption as an underlying uncertainty model. In the context of health and systems monitoring, Kalman filters and their derivatives are often used to track parameter changes over time or to extract knowledge about damage states from noisy measurements. In a more general sense this is also known as sequential updating \parencite{vanik2000bayesian}. As an example, \parencite{ghanem2006health} used an ensemble Kalman filter (EnKF) to estimate damage in a four-story shear building by tracking the changes in system parameters over time. \parencite{erazo2019vibration} showed the application of Kalman filters to structural health monitoring using a decoupling between environmental changes and parameter changes. \parencite{xie2012real} applied an unscented Kalman filter (UKF) to non-linear structural identification. Similarly, \parencite{lei2019novel} proposed an UKF to identify non-linear systems under unknown inputs and \parencite{diaz2023new} used an UKF and error estimation techniques to update a structure subjected to vibrations from a shaking table.

\section{Transport Maps}

A transport map is an invertible deterministic coupling between two probability density functions (PDFs), one being the target density of interest $\pi$ and the other one a user-chosen reference density $\rho$. This coupling allows for the direct sampling and calculations of integrals on the target PDF since calculations can be performed through change of variables \parencite{spantini2018inference}

\begin{equation}
	\int_{Y} f(\bm{y}) \pi(\bm{y}) \text{d}\bm{y} = \int_{X} f(M(\bm{x})) \pi(\bm{x}) \text{d}\bm{x}
\end{equation}
where $M(\cdot) : \mathbb{R}^n \rightarrow \mathbb{R}^n$ is the transport map that transforms samples from the garget density to the reference density. The reference density is usually chosen as standard normal or standard uniform, since this choice makes calculations straight forwards once the map is found \parencite{spantini2018inference}. With the push-forward operator

\begin{equation}
	M_\# \pi = \rho(M(\bm{x})) \cdot |\text{det} \nabla M(\bm{x})|
	\label{eq:ESREL_PushForward}
\end{equation}
and the pull-back operator for the inverted map

\begin{equation}
	M^\# \rho = \pi(M^{-1}(\bm{x})) \cdot | \text{det} \nabla M^{-1} (\bm{x}) |
\end{equation}
a distance for the approximation $M^\#\rho = \pi$ can be calculated with the Kullback-Leibler divergence

\begin{equation}
	\text{D}_{KL}(M^\#\rho || \pi) = \mathbb{E}_\rho \left[\text{log} \frac{M^\#\rho}{\pi}\right]
	\label{eq:ESREL_forwardMap}
\end{equation}
or, due to invertibility of the map,

\begin{equation}
	\text{D}_{KL}(\rho || M_\#\pi) = \mathbb{E}_\rho \left[\text{log} \frac{\rho}{M_\#\pi}\right]
	\label{eq:ESREL_backwardMap}
\end{equation}

The distance can thus either be defined in terms of the reference or the target density. Both options are viable, depending on the use case. If a function for the target density is known, i.e. in Bayesian model updating, where the goal is to approximate the posterior density, the expression in \cref{eq:ESREL_forwardMap} can be used. If only samples from the posterior are known, which is the setting of this paper, the expression in \cref{eq:ESREL_backwardMap} is more useful. Multiple formulations have been proposed for the maps $M$, i.e. polynomials \parencite{marzouk2016sampling}, tensor trains \parencite{dolgov2020approximation} or neural networks \parencite{ardizzone2018analyzing}. In this context, the latter are also referred to as normalizing flows \parencite{kobyzev2020normalizing}. For uniqueness and ease of computation purposes, as well as to fulfil the invertibility criterion, the maps are constructed in a lower-triangular fashion, also known as Knothe-Rosenblatt rearrangement

\begin{equation}
	M(\bm{x}) = \begin{bmatrix}
		M^1(x_1) \\ \vdots \\ M^n(x_1,\cdots,x_n)
	\end{bmatrix}
\end{equation}
where each component $M^i : \mathbb{R}^i \rightarrow \mathbb{R}$ pushes forward the first $i$ components of $\bm{x}$ Moreover, each component is constructed as

\begin{equation}
	M(\bm{a}^i,x_1,\cdots,x_i) = f(\bm{a}^i,x_1,\cdots,x_{i-1}) + \int_{0}^{x_i} g(\partial_i f(\bm{a}^i, x_1, \cdots, x_{i-1},\bar{x}))\text{d}\bar{x}
\end{equation}
with parameterized functions $f(\bm{a}^i,\cdot)$ and a rectifier function $g : \mathbb{R} \rightarrow \mathbb{R}_+$, e.g. the exponential function, that ensures monotonicity in the last argument. For computations we used the MParT framework \parencite{mpart2023}. The functions $f(\bm{a}^i,x_1,\cdots,x_i)$ are defined as linear combinations of Hermite polynomials $\Psi(\cdot)$

\begin{equation}
	f(\bm{a}^i,x_1,\cdots,x_i) = \sum_i \bm{a}_j^i \Psi_{\bm{\alpha}^j} (x_1,\cdots,x_i)
	\label{eq:ESREL_Polynomials}
\end{equation}
with

\begin{equation*}
	\Psi_{\bm{\alpha}^j} (x_1,\cdots,x_i) = \prod_{k=1}^i \Psi_{\alpha_k^j} (x_k)
\end{equation*}

The subscript $\bm{\alpha}^j$ denotes a vector of so-called multi-indices which define the order of the Hermite polynomials. For a more in-depth explanation and derivations the interested reader is referred to \parencite{baptista2023representation} and the references therein.

In order to find the map parameters $\bm{a}$, \cref{eq:ESREL_PushForward} and \cref{eq:ESREL_backwardMap} are written as

\begin{equation}
	\text{D}_{KL}(\rho || M_\#\pi) = \int_{\bm{X}} \Big[\text{log}\, \rho(\bm{x}) - \text{log}\, \rho (M(\bm{a},\bm{x})) - \text{log}\,(|\text{det} \nabla M(\bm{a},\bm{x})|)\Big] \rho(\bm{x}) \text{d}\bm{x}\ .
	\label{eq:ESREL_OptimIntegral}
\end{equation}

A smaller distance indicates a better fit between the approximation and the target PDF. If a number of $N$ samples $\hat{x}$ from the target distance are available, \cref{eq:ESREL_OptimIntegral} can be transformed into a minimization procedure

\begin{equation}
	\underset{\bm{a}}{\arg\min} \frac{1}{N} \sum_i \Big[
	-\text{log}\, \left(\rho(M(\bm{a},\hat{\bm{x}}_i))\right)
	-\text{log}\, \left(|\text{det}\nabla M(\bm{a},\hat{\bm{x}}_i)|\right)
	\Big]\ .
	\label{eq:ESREL_OptimGoal}
\end{equation}

If $\rho$ is chosen as standard normal density, the formulation of \cref{eq:ESREL_OptimIntegral} is equivalent to normalizing the samples $\hat{\bm{x}}$ through the (non-linear) mapping to be distributed according to the standard normal density. The polynomial degree of $\Psi$ in \cref{eq:ESREL_Polynomials} can be changed in order to change the approximation quality, where higher order maps are able to better approximate the target, however, this naturally also increases the computational effort, as the size of $\bm{a}$ increases. \cref{eq:ESREL_OptimGoal} can be solved by any optimization algorithm, especially those that use gradient information have proven to be effective here. The gradient of \cref{eq:ESREL_OptimGoal} is readily available, since all of the expressions are formulated analytically.

\section{Coupling Filter for state estimation}

As a first step we will summarize here the general approach and the formulation for state estimation of a dynamical system to motivate the application to combined state-parameter estimation. This approach was introduced by \parencite{spantini2022coupling}, where more in-depth derivations and proofs can be found. Consider the state-space system with $d$ states and $m$ measurements at some time $k$

\begin{gather}
	\begin{aligned}
		\dot{\bm{x}}_k &= \bm{A}\bm{x}_k + \bm{B}\bm{u}_k \\
		\bm{y}_k &= \bm{C}\bm{x}_k + \bm{w}_k
		\label{eq:ESREL_DynSys}
	\end{aligned}
\end{gather}
where $\bm{x} \in \mathbb{R}^d$ are the states (e.g. positions, velocities...) and their derivatives in time, $\bm{A} \in \mathbb{R}^{d\times d}$ is the state
transition matrix, $\bm{B} \in \mathbb{R}^{d \times d}$ is the input matrix, $\bm{u} \in \mathbb{R}^d$ are inputs to the system, $\bm{y} \in \mathbb{R}^m$ are measurements, $\bm{C} \in \mathbb{R}^{m \times d}$ is the measurement matrix and $\bm{w} \in \mathbb{R}^m$ is additive random noise that corrupts the measurements, i.e. $\bm{w} \sim \pi_W$, where $\pi_W$ is the PDF of $\bm{w}$. Gaussian noise is often used here. The subscript $k$ indicates the discrete time step at time $t_k$. Since entries of the state $\bm{x}$ are only indirectly observable through measurements the goal is to update the system's state $\bm{x}_{k+1}$ at time $t_k$ based on the information obtained through measurements $\bm{y}_{1:k+1}$ up until this point. In most cases, the measurements are also incomplete, i.e. $m < n$. Given some samples $\{\bm{x}_k^1, \cdots, \bm{x}_k^n\}$ at time $t_k$ distributed according to the (assumed known) PDF $\pi_k(\bm{x}_k | \bm{y}_{1:k})$, samples from the prior distribution $\pi_k(\bm{x_{k+1} | \bm{y}_{1:k}})$ at time step $t_{k+1}$ can be obtained by propagating $\bm{x}_k$ forward in time with integration of \cref{eq:ESREL_DynSys}. By constructing a joint probability for measurements and states $\pi_{\bm{Y},\bm{X}}$ and letting the reference PDF $\rho$ be the standard normal density, a map $S(\bm{y},\bm{x}) : \mathbb{R}^{m+d} \rightarrow \mathbb{R}^{m+d}$ can be defined using the push-forward operation from \cref{eq:ESREL_PushForward} as

\begin{equation}
	S_\# \pi_{\bm{Y},\bm{X}} = \rho\ .
	\label{eq:ESREL_FilterPushForward}
\end{equation}

The map $S$ has the structure

\begin{equation}
	S(\bm{y},\bm{x}) =
	\begin{bmatrix}
		S^1(y_1) \\ \vdots \\ S^{m+d}(y_1,\cdots,y_m,x_1,\cdots,x_d)
	\end{bmatrix}\ .
	\label{eq:ESREL_MapStructure}
\end{equation}

Each component $S^i(\bm{y},\bm{x})$ in \cref{eq:ESREL_MapStructure} pushes forward the first $i$ inputs to a marginal distribution of the reference density. Therefore, by definition of the standard normal density, the components are independent and can be computed in parallel. Moreover, since we are only interested in an expression for $\bm{x}$, the first $m$ components can be ignored, giving the final form of $S(\bm{y},\bm{x})$

\begin{equation}
	S(\bm{y},\bm{x}) =
	\begin{bmatrix}
		S^1(\bm{y},x_1) \\ \vdots \\ S^d(\bm{y},x_1,\cdots,x_d)
	\end{bmatrix}
	\label{eq:ESREL_FilterMapStructure}
\end{equation}
where each component is a function $S^i(\bm{y},\bm{x}_{1:i}) : \mathbb{R}^{m+i} \rightarrow \mathbb{R}$.

Samples from $\pi_{\bm{Y},\bm{X}}$ can be obtained by sampling $\bm{y}_{k+1}$ from the noise distribution $\pi_W$ after obtaining prior samples $\hat{\bm{x}}_{k+1}$. Each simulated state $\hat{\bm{x}}_{k+1}$ is thereby augmented by a simulated measurement $\hat{\bm{y}}_{k+1}$ to obtain pairs $(\hat{\bm{y}}_{k+1}^i, \hat{\bm{x}}_{k+1}^i), i \in [1,\cdots,n]$. Note that the posterior is obtained from defining the pairs $(\bm{y}_{k+1}, \bm{x}_{k+1}^i)$ where $\bm{y}_{k+1}$ is the actual measurement at time $t_{k+1}$ and $\bm{x}_{k+1}^i$ are the unknown samples from the posterior density $\pi_{k+1}(\bm{x}_{k+1} | \bm{y}_{1:k+1})$. By definition, these pairs can be pushed to $\rho$ by \cref{eq:ESREL_FilterPushForward}, therefore $\bm{x}_{k+1}^i$ can be obtained after finding the map $S$ by first pushing the pairs $(\hat{\bm{y}}_{k+1}^i, \hat{\bm{x}}_{k+1}^i)$ to the standard normal density, then setting $\hat{\bm{y}}_{k+1} = \bm{y}_{k+1}$ and inverting the map, i.e.

\begin{equation}
	\bm{x}_{k+1} = S^{-1} (\bm{y}_{k+1}, S(\hat{\bm{y}}_{k+1}, \hat{\bm{x}}_{k+1}))
	\label{eq:ESREL_FilterInverseMap}
\end{equation}
where the inverse map $S^{-1}(\bm{y}_{k+1},\cdot)$ is defined by finding the input $\tilde{\bm{x}}$ to $S$ such that

\begin{equation}
	S(\bm{y}_{k+1},\tilde{\bm{x}}) = S(\hat{\bm{y}}_{k+1}, \hat{\bm{x}}_{k+1})
\end{equation}
for the unknown inputs $\bm{y}_{k+1}, \hat{\bm{y}}_{k+1}, \hat{\bm{x}}_{k+1}$. Because of the components' independence, finding the inverse of the map is equivalent to finding the $d$ roots of each scalar function in $S^{-1}$, this can again be done in parallel. The obtained samples $\bm{x}_{k+1}$ from \cref{eq:ESREL_FilterInverseMap} can then be used to update the states for the next time steps in a recursive manner.

\begin{figure}
	\centering
	\includegraphics[width=.8\textwidth]{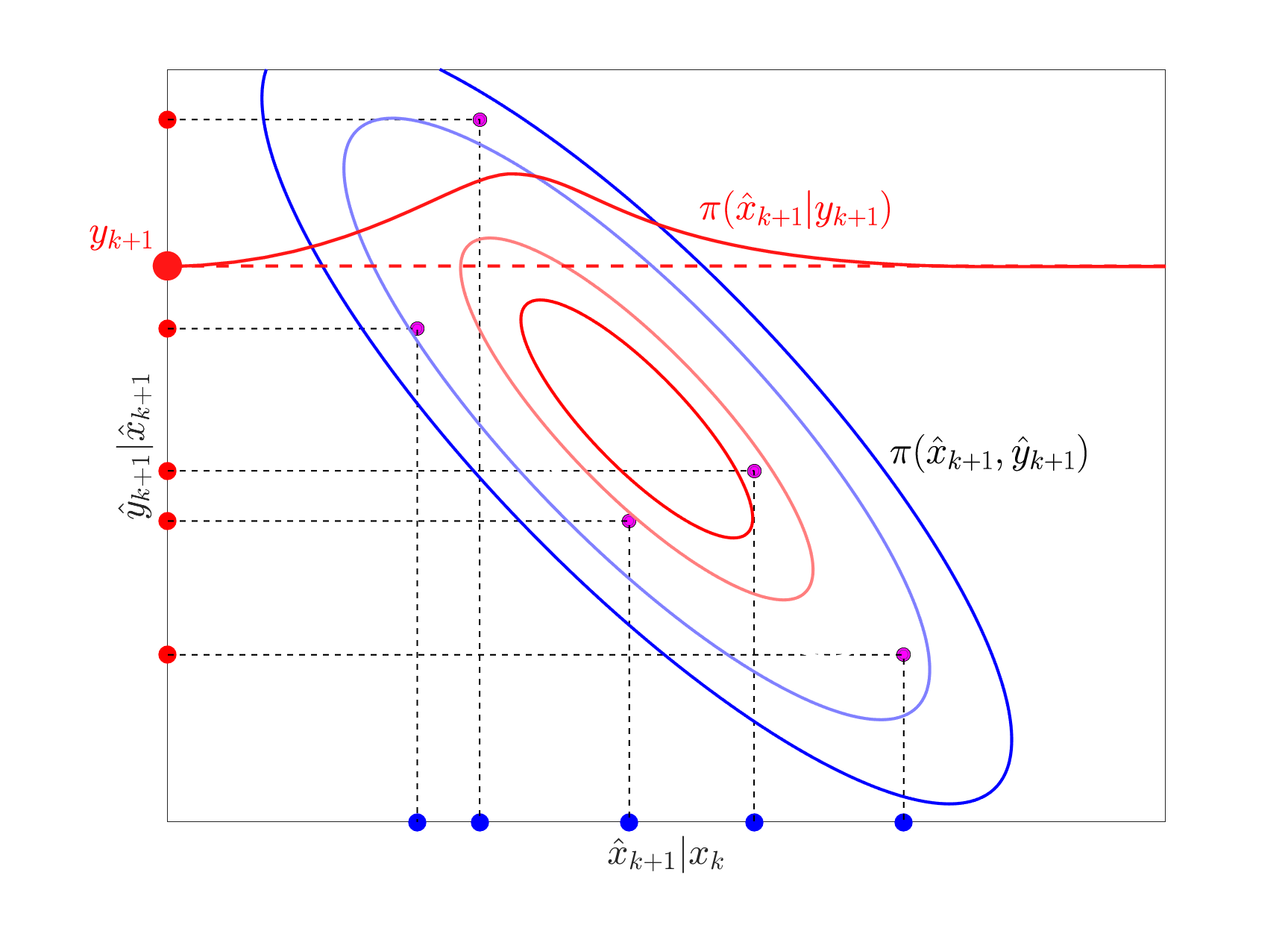}
	\caption{Illustration of the filtering process from time $k$ to time $k+1$ through the joint PDF $\pi_{\bm{Y},\bm{X}}$}
	\label{fig:ESREL_FilteringProblem}
\end{figure}

An illustration for the filtering process between two time steps is shown in \cref{fig:ESREL_FilteringProblem}. The blue samples represent the propagated states from time $k$ to time $k+1$, i.e. they are sampled according to $\pi_k(\bm{x}_{k+1} | \bm{x}_k)$. Red samples indicate the likelihood samples that were drawn from $\pi_k(\hat{\bm{y}}_{k+1} | \hat{\bm{x}}_{k+1})$. Together, these represent samples (magenta) from the joint PDF $\pi_{\bm{Y},\bm{X}}$ (indicated as contour plot), which can be approximated by the transport map formulation to an arbitrary degree. After obtaining this approximation it can be used to condition the joint PDF on a measurement $\bm{y}_{k+1}$ by simply inverting the map, thus reaching the desired posterior PDF $\pi(\hat{\bm{x}}_{k+1} | \bm{y}_{k+1})$ (red line in \cref{fig:ESREL_FilteringProblem}).

The advantage of using transport maps to approximate the joint density is that there is virtually no assumption on the structure of the involved probability functions, especially on $\pi_{\bm{Y},\bm{X}}$. As long as the map parametrization is rich enough to describe the involved densities, neither the state transition nor the noise needs to follow Gaussian or linear assumptions, as is the case with many other filtering algorithms.

\section{Joint state-parameter estimation using the coupling filter}

Many dynamical systems are dependent on some parameters that define their behavior in time. Let $\ftheta \in \mathbb{R}^{d_{\ftheta}}$ be a vector of parameters for the system described in \cref{eq:ESREL_DynSys}, then it can be rewritten as

\begin{gather}
	\begin{aligned}
		\dot{\bm{x}}_k &= \bm{A}(\ftheta)\bm{x}_k + \bm{B}(\ftheta)\bm{u}_k \\
		\bm{y}_k &= \bm{C}(\ftheta)\bm{x}_k + \bm{w}_k
		\label{eq:ESREL_ParameterUpdating}
	\end{aligned}
\end{gather}
where the system's matrices are now dependent on $\ftheta$. If $\ftheta$ is unknown and subject to identification, the state vector $\bm{x}$ can be augmented by simply adding $\ftheta$ as entries in $\bm{x}$, i.e.

\begin{equation}
	\bm{x} =
	\begin{bmatrix}
		x_1 & \cdots & x_d & \theta_1 & \cdots & \theta_{d_{\ftheta}}
	\end{bmatrix}^\top \ .
\end{equation}

Estimating both states and parameters together follows the same procedure as described in the previous chapter, however the orders of magnitude in each dimension of $\bm{x}$ can vary drastically which makes solving the optimization
procedure in \cref{eq:ESREL_OptimGoal} inefficient. We therefore introduce an additional regularization step before building the map $S$, which decreases the computational effort without adding too much complexity. Since the standard normal density is used as a reference, the prior pairs $\hat{\bm{\gamma}}_{k+1} = (\hat{\bm{y}}_{k+1}, \hat{\bm{x}}_{k+1})$ and $\tilde{\bm{\gamma}}_{k+1} = (\bm{y}_{k+1}, \hat{\bm{x}}_{k+1})$ can be normalized to a zero-mean distribution with unit covariance by shifting and scaling the samples with

\begin{gather}
	\begin{aligned}
		\hat{\bm{\chi}}_{k+1} &= \bm{\Sigma}_{k+1}^{-1} \cdot (\hat{\bm{\gamma}}_{k+1} - \bar{\hat{\bm{\gamma}}}_{k+1}) \\
		\tilde{\bm{\chi}}_{k+1} &= \bm{\Sigma}_{k+1}^{-1} \cdot (\tilde{\bm{\gamma}}_{k+1} - \bar{\hat{\bm{\gamma}}}_{k+1})
		\label{eq:ESREL_NormalizedPairs}
	\end{aligned}
\end{gather}
where $\hat{\bm{\chi}}_{k+1}$ and $\tilde{\bm{\chi}}_{k+1}$ denote the normalized samples, $\bm{\Sigma}_{k+1}^{-1}$ is the inverse lower-triangular Cholesky factor of the covariance matrix of the prior samples and $\bar{\hat{\bm{\gamma}}}_{k+1}$ is the mean of the prior samples. Note that the pair $\hat{\bm{\gamma}}_{k+1}$ contains the simulated measurements, while the pair $\tilde{\bm{\gamma}}_{k+1}$ contains the actual measurements. Both are normalized using the same mean and covariance. The resulting normalized samples more closely resemble the target distribution which decreases the needed effort to find the map, since the distance to the target is smaller. Moreover, the transformation in \cref{eq:ESREL_NormalizedPairs} is linear and easily available, because it only involves calculating the mean and covariance of the prior samples. The objective now is to find normalized posterior samples $\bm{\gamma}_{k+1}$, which can be transformed back to the non-normalized posterior pairs $\bm{\gamma}_{k+1} = (\bm{y}_{k+1},\bm{x}_{k+1})$ by

\begin{equation}
	\bm{\gamma}_{k+1} = \bm{\Sigma}_{k+1} \cdot (\tilde{\bm{\chi}}_{k+1,1:m}, \bm{\chi}_{k+1}) + \bar{\hat{\bm{\gamma}}}_{k+1}
	\label{eq:ESREL_BackTransform}
\end{equation}
where $\bm{\chi}_{k+1}$ is the output of the map inversion from \cref{eq:ESREL_FilterInverseMap}. Since $\bm{\chi}_{k+1} \in \mathbb{R}^d$ and $\bm{\gamma}_{k+1} \in \mathbb{R}^{m+d}$ the first $m$ normalized measurements from $\tilde{\bm{\chi}}_{k+1}$ are used in the back-transformation in \cref{eq:ESREL_BackTransform}. The full procedure to estimate states and parameters is summarized below:

\begin{itemize}
	\item at time $t_{k+1}$ propagate $\bm{x}_k$ to $\bm{x}_{k+1}$ through the (non-linear) system [\cref{eq:ESREL_ParameterUpdating}]
	\item sample pairs $\hat{\bm{\gamma}}_{k+1}$ from $\pi_{\bm{Y},\bm{X}}$
	\item normalize obtained samples to obtain $\hat{\bm{\chi}}_{k+1} and \tilde{\bm{\chi}}_{k+1}$ [\cref{eq:ESREL_NormalizedPairs}]
	\item build and optimize $S(\hat{\bm{\chi}}_{k+1})$ [\cref{eq:ESREL_FilterMapStructure} and \cref{eq:ESREL_OptimGoal}]
	\item calculate inverted map output fixed on the measurements [\cref{eq:ESREL_FilterInverseMap}]
	\item calculate non-normalized posterior samples $\bm{\gamma}_{k+1}$ [\cref{eq:ESREL_BackTransform}]
\end{itemize}

\subsection{Likelihood oversampling}

In order to enhance the efficiency of the filter we propose to oversample the likelihood, i.e. draw more than one sample per propagated state. This increases the information content in the approximation of the joint distribution $\pi_{\bm{Y},\bm{X}}$ and makes it more robust, since the maps are able to better approximate the joint PDF. While drawing samples from the likelihood is computationally inexpensive, this slightly increases the time it takes to compute the transport maps since more samples need to be evaluated. This is however favorable over simulating more system states since this would require more evaluations of the system's equations, which usually is more time-consuming.

\section{Numerical example}

\begin{figure}
	\centering
	\includegraphics[width=\textwidth]{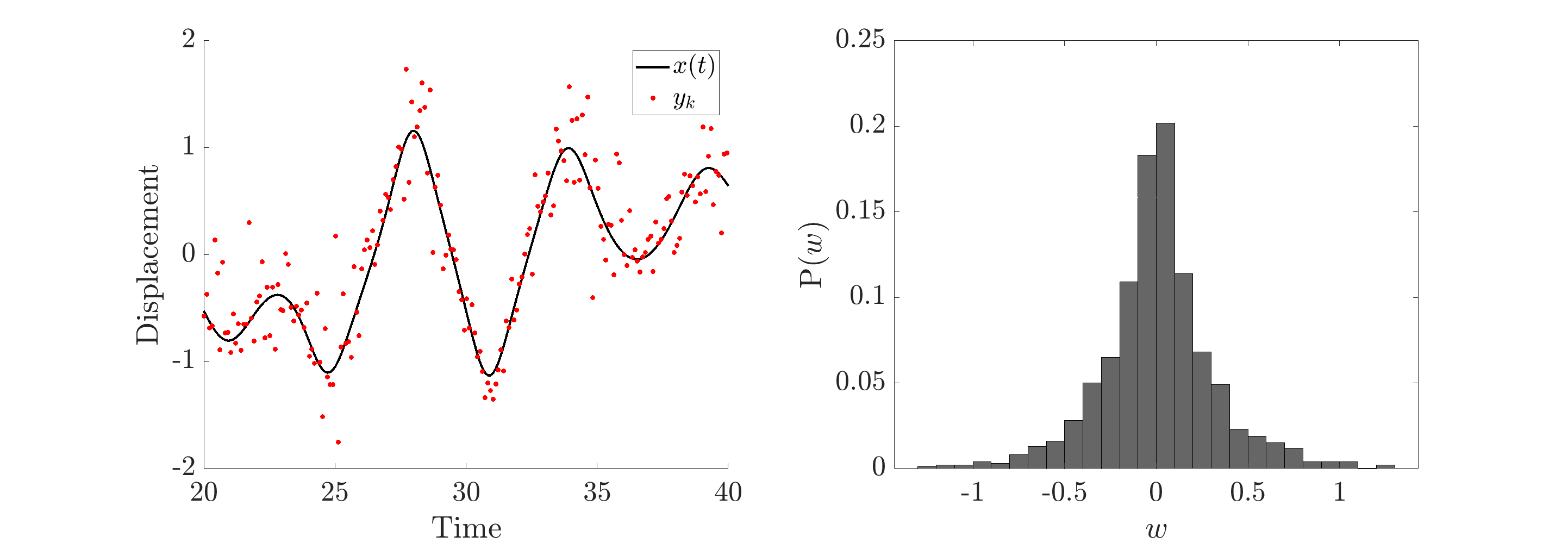}
	\caption{Example for system behavior and noise}
	\label{fig:ESREL_Systembehavior}
\end{figure}

In order to show the applicability of the coupling filter to non-linear state-parameter estimation, we will show the procedure here on a Duffing oscillator. The equation of motion is

\begin{equation}
	\ddot{x} + \delta \dot{x} + \alpha x + \beta x^3 = \kappa\, \text{cos} (\omega_0 t)
	\label{eq:ESREL_Duffing}
\end{equation}
with scalar states $\ddot{x}, \dot{x}, x$ that denote acceleration, velocity and position, linear stiffness $\alpha$, linear damping $\delta$ and non-linear stiffness $\beta$. In this example we use a forced Duffing oscillator with force amplitude $\kappa$ and frequency $\omega_0$. \cref{eq:ESREL_Duffing} can be written in state-space

\begin{equation}
	\overbrace{\begin{bmatrix}
		\dot{x} \\ \ddot{x}
	\end{bmatrix}}^{\dot{\bm{x}}}
	=
	\overbrace{
	\begin{bmatrix}
		0 & 1 \\ -\alpha & -\delta
	\end{bmatrix}}^{\bm{A}}
	\overbrace{
		\begin{bmatrix}
			x \\ \dot{x}
	\end{bmatrix}}^{\bm{x}}
	+
	\overbrace{
	\begin{bmatrix}
		0 & 0 \\ 0 & 1
	\end{bmatrix}
	}^{\bm{B}}
	\overbrace{
	\begin{bmatrix}
		0 \\ \kappa\, \text{cos}(\omega_0 t)
	\end{bmatrix}
	}^{\bm{u}}
	+
	\overbrace{
	\begin{bmatrix}
		0 & 0 \\ -\beta & 0
	\end{bmatrix}	
	}^{\bm{A}_{nl}}
	\begin{bmatrix}
		x \\ \dot{x}
	\end{bmatrix}^3
\end{equation}
with measurements

\begin{equation}
	\begin{bmatrix}
		y \\ \dot{y}
	\end{bmatrix}
	=
	\bm{C}\bm{x} + w
\end{equation}
For estimation of the parameters, $\bm{x}$ is augmented as $\bm{x} = [x, \dot{x}, \alpha, \delta, \beta]^\top$ and $C = [1\ 0\ 0\ 0\ 0]$, i.e. only the position is measured directly. The noise $w$ is chosen to follow zero-mean Laplace distribution, since this exhibits much longer tails than a Gaussian distribution. The equation for the Laplace distribution (omitting the mean value) is

\begin{equation}
	\pi_w (x) = \frac{1}{2\sigma_w} \text{exp} (- \frac{|x|}{\sigma_w})
\end{equation}
with $\sigma_w = 0.3^2$. We use here $\alpha = -1$, $\delta = 0.3$ and $\beta = 2$ for the system parameters to be identified, as well as $\kappa = 0.5$ and $\omega_0 = 1.2$ for the input parameters. To show the effect of likelihood oversampling we calculated results for a single and three likelihood samples per state. As initial conditions we set the position and velocity to zero and set the initial guesses for the parameters to be twice the value of the actual parameters to verify that the filter is able to deal with non-optimal initialization. Further, we use a sample size of 20 systems, with the oversampling described as above this leads to 20 and 60 samples to approximate $\pi_{\bm{Y},\bm{X}}$, respectively. We simulate the system in $T = [0,100]$ with time step size $\Delta t = 0.1$. An example for the system behavior and the histogram of the noise is shown in \cref{fig:ESREL_Systembehavior}. The figure shows the system displacement (black line) from $t = 20$ to $t = 40$ and the measurements (red dots) on the left. The right diagram shows the relative occurrence of noise values over the full simulated time.

\begin{figure}
	\centering
	\includegraphics[width=\textwidth]{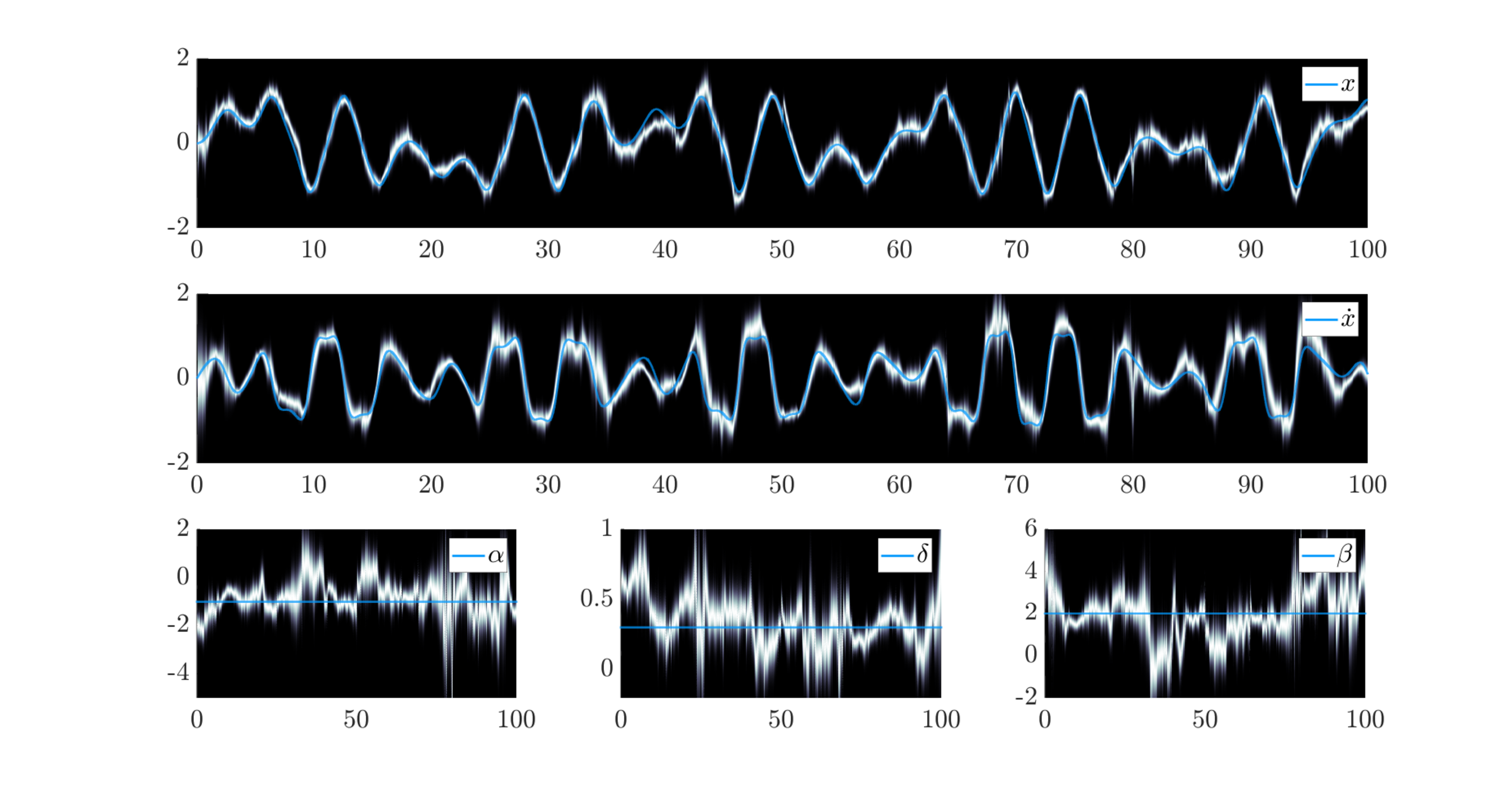}
	\caption{Results for no oversampling for updating of Duffing-oscillator, figures show the PDF of the states and parameters.}
	\label{fig:ESREL_Result1versampling}
\end{figure}

\begin{figure}
	\centering
	\includegraphics[width=\textwidth]{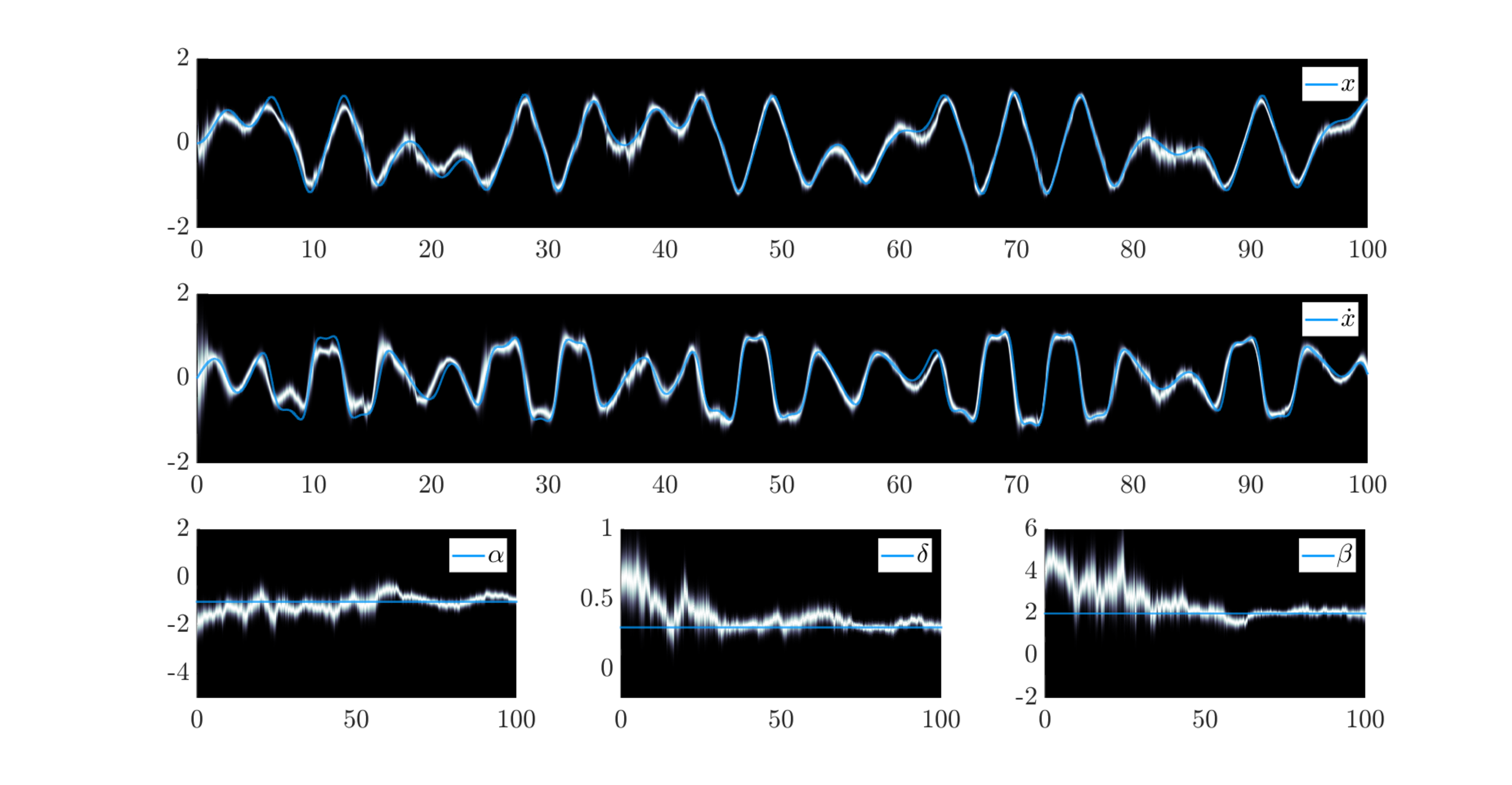}
	\caption{Results for an oversampling factor of 3 for updating of Duffing-oscillator, figures show the PDF of the states and parameters.}
	\label{fig:ESREL_Result3versampling}
\end{figure}

Results of the identification are shown in \cref{fig:ESREL_Result1versampling} and \cref{fig:ESREL_Result3versampling}, where the top two graphs show the results for position $x$ and velocity $\dot{x}$ and the bottom row shows the estimations for the three parameters $\alpha$, $\delta$ and $\beta$. True values are given as a blue line, all graphs show the estimated posterior PDFs in time shaded in white. In the case without oversampling (\cref{fig:ESREL_Result1versampling}) it can be observed that the state and velocity are estimated quite well, although with greater variance than if oversampling is used (\cref{fig:ESREL_Result3versampling}). Especially the uncertainty in the non-linear stiffness parameter $\beta$ is very large and it deviates quite far from the true value around $t = 40$. A similar situation can be observed for the linear stiffness $\alpha$. The damping $\delta$ is estimated with less variance. The estimations for $x$ and $\dot{x}$ also deviate from the
true state, however they do not become unstable, which could be a concern in this example since the system behaves slightly chaotic.

\begin{figure}
	\centering
	\includegraphics[width=.8\textwidth]{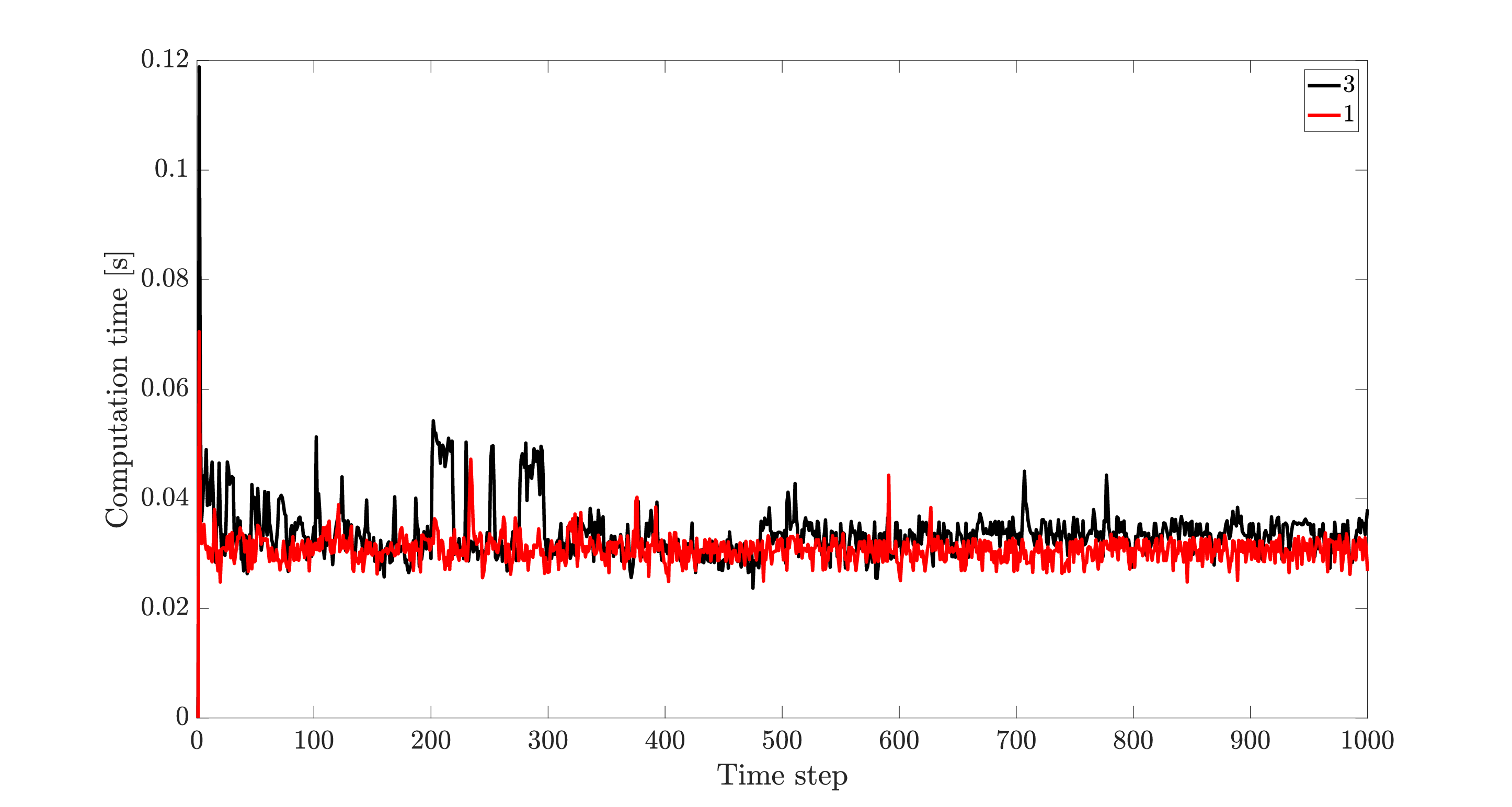}
	\caption{Map computation time at each time step for oversampling factors of 1 (red) and 3 (black)}
	\label{fig:ESREL_MapTime}
\end{figure}

Oversampling the likelihood with three samples per propagated state (\cref{fig:ESREL_Result3versampling}) overall results in less variance in
the estimations. Position and velocity closely follow the true values and the three parameters converge to the prescribed values. From this example it is evident that an oversampling of the likelihood provides more information about $\pi_{\bm{Y},\bm{X}}$ without drastically increasing the computational burden. To illustrate this further, \cref{fig:ESREL_MapTime} shows the computation time for the maps at each time step, with the no oversampling case in red and the oversampling in black. Note that these times do not include the time for the state propagation from step $k$ to step $k+1$. Since we used 20 samples from the system in both cases the time to propagate the system is identical. It can be seen that, on average, the time for the oversampled maps are longer than the times for maps without oversampling. The overall computation times, including model propagation, were 36.55 s without oversampling and 39.60 s with oversampling. Also note that the computation time in each time step roughly stays the same, which is an important criterion for real-time state-parameter estimation.

The difference between the two cases depends on multiple factors like the map dimensionality, map order, the oversampling factor or the topology of $\pi_{\bm{Y},\bm{X}}$. Increasing the map dimensionality, i.e. increasing the number of estimated states and parameters, or increasing the map order increase the map complexity and therefore the optimization time. If oversampling is used, the maps need to be evaluated more often. The longer computation times in the maps themselves then lead to a larger difference in the evaluation times when the likelihood is oversampled. The same argument is true for the oversampling factor, i.e. more samples from the likelihood lead to more computation time, as we showed in the example in this paper. Generally, a balance should be found between the number of simulated system states $\hat{x}_{k+1}$ and the number of samples from the likelihood $\hat{y}_{k+1}$. \Cref{fig:ESREL_FilteringProblem} gives some intuition for this process. In this two-dimensional example a larger number of samples $\hat{x}_{k+1}$ directly correspond to a larger number of samples from $\pi_{\bm{Y},\bm{X}}$ (if one likelihood sample per propagated state is drawn, i.e. no "undersampling" is performed) which adds information about the uncertainty of parameters and measurement noise at the same time. On the other hand, if the likelihood is oversampled, only information about the noise is added. Oversampling thus increases the robustness against measurement noise to some extent but should also not be exaggerated. At some point there is little gain from increasing the number of samples from the likelihood, however, this would still increase the computation time.

\section{Conclusion}

In this contribution a coupling filter for joint state-parameter estimation for non-linear dynamical systems based on transport maps was introduced and evaluated. The filter works by approximating the joint density of states and measurements with so-called transport maps, which give an analytic formulation of the density. This analytic formulation can then be used to condition the joint PDF at any assimilation time step on a measurement, giving an approximation for the posterior density. In the used transport map framework this operation is done by simply inverting polynomial expressions which make the process very efficient. Moreover, since there are no assumptions on the noise distributions or the model, non-Gaussian and non-linear settings can be treated in a straight-forward manner. The presented filter performed well for the state and parameter estimation of a Duffing oscillator with non- optimal initial conditions under non-Gaussian noise. We also evaluated likelihood oversampling, which can help to increase the available information about the measurement noise and make the filter more robust against it. We found that oversampling the likelihood function decreases the overall variance in the estimations.

\section*{Acknowledgement}

Funding by the Deutsche Forschungsgemeinschaft (DFG, German Research Foundation) for the GRK2657 (grant reference number 433082294) is greatly appreciated.

\printbibliography[title=References]

\end{document}